\documentclass[a4paper,12pt]{article}
\pdfoutput=1
\usepackage{amssymb,amsmath,bm}
\usepackage{epigraph}
\usepackage{color}
\usepackage{slashed}
\usepackage{cite}
\usepackage[latin1]{inputenc}
\usepackage{amsfonts}
\usepackage{lscape}
\usepackage{amsthm}
\usepackage{booktabs}
\usepackage{array}
\usepackage{rotating}
\usepackage{float}
\usepackage{multirow}
\usepackage{graphicx,rotating,amsmath,multirow,xcolor,colortbl,xcolor}
\usepackage{hyperref,pdflscape,slashed}

\font\tenrsfs=rsfs10 at 12pt
\font\sevenrsfs=rsfs7
\font\fiversfs=rsfs5
\newfam\rsfsfam
\textfont\rsfsfam=\tenrsfs
\scriptfont\rsfsfam=\sevenrsfs
\scriptscriptfont\rsfsfam=\fiversfs
\def\mathscr#1{{\fam\rsfsfam\relax#1}}

\usepackage[small,bf]{caption}
\newcommand{\bea}{\begin{eqnarray}}
\newcommand{\eea}{\end{eqnarray}}
\newcommand{\beq}{\begin{equation}}
\newcommand{\eeq}{\end{equation}}

\begin{document}

{\hfill CERN-TH-2017-205}

\vspace{1cm}

\begin{center}
\boldmath

{\textbf{\LARGE The Dawn of the Post-Naturalness Era}}

\unboldmath

\bigskip

\vspace{0.4 truecm}

{\bf Gian Francesco Giudice} 
 \\[5mm]

{\it CERN, Theoretical Physics Department, Geneva, Switzerland}\\[2mm]

\vspace{0.8cm}

{\bf Abstract }

\vspace{0.2cm}
\begin{quote}
In an imaginary conversation with Guido Altarelli, I express my views on the status of particle physics beyond the Standard Model and its future prospects.
\end{quote}

\vspace{0.8cm}
Contribution to the volume {\it ``From My Vast Repertoire" -- The Legacy of Guido Altarelli}.
\end{center}

\vspace{1.5cm}

\section{A Master and a Friend}
\label{sec1}
\setlength\epigraphwidth{.7\textwidth}
\setlength\epigraphrule{0pt}
\renewcommand{\textflush}{flushright}
\epigraph{\it Honour a king in his own land; honour a wise man everywhere.}
{---~~Tibetan proverb~\cite{tibet}}

Guido Altarelli was an extraordinary theoretical physicist. Not only was Guido one of the heroes of the Standard Model, but he incarnated the very essence of that theory: a perfect synthesis of pure elegance and brilliance. With his unique charisma, he had a great influence on CERN and contributed much in promoting the role of theoretical physics in the life of the laboratory. With the right mixture of vision, authority, and practical common sense, he led the Theory Division from 2000 to 2004. I have always admired Guido for his brilliance, humour, knowledge, leadership, and intellectual integrity. I learned much from his qualities and his example is a precious legacy for me and for all of his colleagues. 

Guido had a very pragmatic attitude towards scientific theories. He was not attracted by elaborate mathematical constructions, but wanted to understand the essence behind the formalism and get straight to the concept. In physics, I would define him as a conservative revolutionary or as an optimistic skeptical. One episode illustrates the meaning of this definition. 

In October 2003, I went to Fermilab for a Workshop on Future Hadron Colliders. One morning Guido, Fabiola Gianotti, and I went for breakfast in a coffee shop in Naperville. While the blueberry muffins were markedly forgettable, our conversation was memorable. Although at the time ATLAS was only an empty cavern, the LHC was already very present in everyone's mind and so Fabiola started to ask what should we expect the LHC to find: supersymmetry, technicolor, extra dimensions, little Higgs? Nothing, said Guido dispassionately, nothing other than the Higgs. Guido had a natural skepticism towards complicated model building or elaborate constructions whose only rationale seemed to be their ability of ignoring what LEP was telling us. But never did his skepticism turn into a bleak view of the future. He kept a sober attitude towards the latest fashions in new-physics theories, but had a sincere interest in them. Prejudices were not hampering his scientific curiosity. His natural optimism was not incompatible with his rational skepticism. This wise balance is a useful lesson to learn for any scientist who experiences the rollercoaster of excitements and disappointments that characterise research at the frontier of knowledge. I will not forget our breakfast in Naperville.

Some of the views expressed during our breakfast conversation reverberated in the summary talk that he gave at the Fermilab workshop. With his inimitable theatrical gestures and distinctive Roman accent, he concluded his talk by saying: ``I think that the case for fundamental physics (particle physics and cosmology) is strong beyond fashion and that we should defend it proudly and confidently."~\cite{altwork} These words still echo in my mind. I shared with Guido this point of view in 2003 and I still passionately subscribe to it today.
  
Guido had a significant influence on my personal scientific growth. We were not regular collaborators. We coauthored only three scientific papers, but one of them~\cite{Altarelli:1997ce} was remembered by Graham Ross as ``the most {\it staffed} paper I have ever seen from CERN." Having four staff members among the authors of the same research paper was peculiar for the CERN Theory Division whose traditional hallmark has always been a respectable (and healthy) dose of individualism. 

In spite of the limited collaboration, we had regular physics conversations. Sometimes he was coming into my office, asking me about new trends in physics Beyond the Standard Model; often I was visiting his office showing him some new results and asking him for advice. I regarded his opinions highly and our conversations contributed greatly to my scientific development. I consider him as one of my masters, but I am honoured to say that he treated me like a friend.

My contribution to this volume has the same style as past conversations with Guido. I will expand freely on views about the status of our field and its future prospects, with no intention to be systematic or complete. My only goal is to have one last conversation with Guido, telling him how I see the field evolving, with no presumption of knowing what is really going to happen in the future. (Pragmatic realism in scientific matters is something I learned from Guido.) My point will be that we live in times of great uncertainties -- the best moments for scientific revolutions to happen -- so most of what I say today is likely to become obsolete very soon. And I sincerely hope so, because I am eagerly waiting for radical new ideas to disrupt our present beliefs. Guido, too, would like that to happen. 

\section{The Naturalness Era}
\label{sec2}
\setlength\epigraphwidth{.5\textwidth}
\setlength\epigraphrule{0pt}
\renewcommand{\textflush}{flushright}
\epigraph{\it Naturalness is the seal of genius.}
{---~~Fran\c{c}ois-Joachim de Pierre de Bernis~\cite{nat}}

If one has to summarise in one word what drove the efforts in physics beyond the Standard Model of the last several decades, the answer is {\it naturalness}. (For a non-technical introduction to the naturalness problem and for my views on the subject, see \cite{Giudice:2008bi,Giudice:2013nak}.) Like it or not, naturalness has been the leitmotiv that has accompanied and motivated most of the attempts to incorporate the Standard Model into a bigger framework at the weak scale. There are good reasons for that. First, no mechanism for electroweak breaking can be viewed as complete without an explanation of the size of the Fermi scale relative to other scales that enter particle physics. Second, naturalness has shown to be a good guiding principle to infer the cutoff scale of a theory in a variety of cases: from the emergence of QED beyond energies of the order of the electron mass, to the presence of hadronic resonances in the GeV range, or the existence of the charm quark \cite{Giudice:2008bi}. This success is related to the fact that naturalness is an inescapable consequence of the ingredients generally used to construct effective field theories. Finally, there is a much more practical reason for the emphasis on naturalness in research beyond the Standard Model. When applied to the Higgs mechanism, naturalness indicates the need for new physics at a scale within the reach of the present generation of collider experiments. This combination of favourable factors has made naturalness the bull's eye of research in physics beyond the Standard Model.

The effort of understanding the role of naturalness in electroweak physics was rewarded with some of the most extraordinary ideas in theoretical physics: ideas with the ambition to redesign the very structure of spacetime and reinvent the pattern of fundamental forces. Technicolor, supersymmetry, extra dimensions, composite Higgs: all these theories proposed a new interpretation of physical reality and opened new avenues in the exploration of the particle world. These avenues were enthusiastically pursued by experimental research and the LHC is the critical step in this journey. 

The LHC delivered an integrated luminosity of about 40~fb$^{-1}$ at $\sqrt{s} = 13$~TeV by the end of 2016 and is rapidly accumulating more data as Run 2 unfolds. This result may look like only a small step in a project that aims at reaching a total of 3000~fb$^{-1}$ by the end of the high-luminosity phase (around 2035). But this small step has given us already a wealth of information, radically influencing the directions of research and the way theorists think of the microworld. 

The 2012 discovery of the Higgs boson~\cite{Aad:2012tfa} was a crucial milestone that revealed how the Standard Model completes the spontaneous breaking of the electroweak symmetry at short distances. The simplest possible UV completion, with a single weakly-interacting scalar particle, received striking experimental confirmation. At the same time no evidence was found of dynamics able to cure the sickness of the Higgs with respect to naturalness. By the summer of 2017, the limits on new particles have reached levels of tension with naturalness of more than the percent: in the simplest supersymmetric setups (which correspond to the most aggressive bounds) the gluino mass limits are about 2 TeV~\cite{donof}, while scalar or fermion top partners are constrained up to 1 TeV~\cite{donof}. Although more data is needed to cast the final verdict on the role of naturalness in electroweak physics, the present data have already had a decisive role in ruling out many model-building possibilities and in reorienting the thinking of most theorists.

A celebrated implication of weak-scale solutions to the Higgs naturalness problem is known as the {\it WIMP miracle}. With apparently innocuous assumptions, often automatically satisfied in realistic models, one finds the intriguing coincidence that the same ingredients needed to cure an unnatural Higgs also lead to a population of particles in today's universe with a density compatible with dark matter observations. The result is striking. Two seemingly unrelated problems -- the Higgs unnaturalness (rooted in the quantum structure of the particle world at distances of $10^{-20}$ meters and below) and the nature of dark matter (observed from galactic distances of $10^{20}$ meters to the largest scales in the universe) -- are linked to the same physical agent. This is the dream of every physicist: a unified and conceptually simple explanation of completely different phenomena.

The conceptual success of the WIMP miracle has ignited a robust experimental program for the search of weak-scale dark matter particles. So far, no evidence for the existence of such particles has emerged and the results translate into strong constraints on models. 
In many realistic constructions, direct detection experiments rule out the case of weak-scale thermal relics, if the dark-matter particle has spin-independent couplings with ordinary matter in the non-relativistic regime. 
The case of spin-dependent interactions is more unconstrained, and here indirect detection gives significant (but mostly model dependent) bounds.

\section{A Krisis?}
\label{sec3}
\setlength\epigraphwidth{.55\textwidth}
\setlength\epigraphrule{0pt}
\renewcommand{\textflush}{flushleft}
\epigraph{\it If a problem is fixable, then there is no need to worry.
If it's not fixable, then there is no help in worrying.}
{---~~His Holiness the 14$^{\rm th}$ Dalai Lama~\cite{dalai}}

In his famous essay {\it The Structure of Scientific Revolutions}~\cite{kuhn}, the science historian Thomas Kuhn identifies a pattern in the development of scientific theories that is common to all revolutions in science. By freely reinterpreting (and simplifying) Kuhn's structure, I can distinguish three phases in the process. The {\it phase of discovery} is when new conceptual breakthroughs and experimental results lead to the emergence of a new theory that departs from old paradigms. This is followed by a {\it phase of consolidation}, in which the theory is understood at a much deeper level and confirmed by precise measurements. This process has the effect of transforming the new theory into the established paradigm of {\it normal science}. Inescapably, this is superseded by a {\it phase of crisis}, in which the normal theory can no longer address new conceptual questions or explain experimental data. This phase is characterised by the search for new paradigms and marked by periods of confusion and frustration. Finally a {\it paradigm shift} occurs, which results in a radical departure from normal science, thus activating a new phase of discovery and marking the beginning of a new cycle.

We can recognise this pattern in the development of both particle physics and cosmology. The Standard Model of particle physics went through its phase of discovery from the late 60's to the early 80's with the rise of gauge theories -- from the use of spontaneous symmetry breaking in weak interactions (1967) to asymptotic freedom in strong interactions (1973) -- and with the experimental discoveries of charm (1974), tau (1975), bottom (1977), W and Z (1983). The phase of consolidation was characterised, on the theory side, by a careful exploration of the Standard Model at the quantum level and, on the experimental side, by the confirmation of the gauge doctrine as the ruling principle in the world of particles through the LEP program and by the discoveries of the top (1995) and Higgs (2012). 

The discovery of the cosmic microwave background, the quantitative predictions in nucleosynthesis, the emergence of dark matter and dark energy, the understanding of initial conditions and structure formation in terms of an inflationary theory have marked the birth of cosmology as a science and characterised its discovery phase. This was followed by an extensive program in observational cosmology (precision studies of the cosmic microwave background, large-scale surveys, gravitational lensing, telescopes from radio to gamma rays), which proclaimed $\Lambda$CDM as the standard model of cosmology and which can be identified with the phase of consolidation.

There are many indications that, following the recursive pattern of scientific revolutions, we are now witnessing the beginning of the phase of crisis. The lack of new physics in the initial stages of the LHC project is putting into question the logic of naturalness when applied to the Higgs; the absence of a positive detection in dark matter experiments is casting doubts about nature taking advantage of the WIMP miracle. We are not simply confronted with experimental data excluding a model or a class of models. We are confronted with the need to reconsider the guiding principles that have been used for decades to address the most fundamental questions about the physical world. These are symptoms of a phase of crisis.

The Standard Model of particle physics is a superb monument attesting to the inner beauty of nature and the power of human logical deduction. It is astounding how natural phenomena, in all their complexity, can be summarised by a single principle -- the {\it gauge} principle -- and described by a compact set of equations. And it is equally astounding how humans have been able to crack this secret. Along this path, the synthesis of general relativity with the physical laws derived in the microworld has led to the $\Lambda$CDM model, which can successfully describe a huge array of cosmological observations, the present large-scale structure of the universe and its early history, in terms of a handful of parameters. This is today's consolidated {\it normal science}. 

And yet, this superb monument of knowledge is insufficient to address some fundamental questions. The Standard Model is incapable of shedding light on the dynamics underlying electroweak symmetry breaking or explaining the structure of quarks, leptons, and their mass pattern at a fundamental level. The theory of inflation, in spite of its stunning conceptual successes, could not be linked univocally with a unified theory of particle physics. Moreover, the ubiquitous phenomenon of eternal inflation has changed the perspective on the outcome of an inflationary universe and its properties. We have plausible explanations for the cosmic baryon asymmetry, but we lack any conclusive empirical confirmation. The nature of dark matter is still unknown. The observed value of the cosmological constant is hard to reconcile with the rules of effective field theory, and quantum gravity is still beyond our grasp.

None of these problems are new, and theoreticians have been tackling them for decades. What is changing is the feeling that the paradigm that so successfully led to the Standard Model may not be the right tool to make further progress. There is a widespread sensation that the organising principles based on symmetry and separation of scales, which follow from an effective quantum field theory approach, in spite of their triumphs, must be superseded by new organising principles. Physicists are in search for new conceptual paradigms, which is another symptom of a phase of crisis.

\subsubsection*{The privilege of experiencing krisis}

Today the word crisis has a sinister connotation, suggesting an approaching downfall, a moment of difficulty or danger. This is not the original meaning of the word. Crisis comes from the Greek {\it krisis}, which means ``decisive moment", ``turning point", and was especially used in a medical context by Hippocrates and Galen as the end of a disease. This is the meaning I will refer to and, to stress the idea, I will use the spelling krisis.

Krisis means the opportunity to look at a problem with new eyes; it is a moment of change, a discontinuity between past and future. Krisis does not mean a decline of ideas, but the search for a paradigm change. 

Each phase of the Kuhnian cycle is a necessary step for a scientific revolution to happen and an exciting moment for the scientists who live through it. The phase of discovery is the moment of eureka. The phase of consolidation is the moment of maturity. The phase of krisis is the moment of opportunity.\footnote{In a 1959 speech John F.~Kennedy stated: ``In the Chinese language, the word `crisis' is composed of two characters, one representing danger and the other opportunity."~\cite{kennedy1}  Although perfectly befitting in my context, Kennedy's assertion is unfortunately incorrect as it is based on an etymological misinterpretation of the word {\it w\=ei-j\=\i}~\cite{zimmer}. In 1963, Kennedy had another unfortunate encounter with foreign languages, when he misquoted Dante as saying: ``The hottest places in hell are reserved for those who in time of moral crisis preserve their neutrality"~\cite{kennedy2}, as a result of a dubious translation of the verses: ``Questo misero modo tegnon l'anime triste di coloro che visser sanza 'nfamia e sanza lodo."~\cite{dante} Once again, notwithstanding the literary fallacy, Kennedy's words serve my purpose. I concur with him in urging all physicists to give up neutrality and take a stand on the issue of naturalness in these times of scientific crisis.}
 A young researcher entering the field can find satisfaction in each of these phases: witnessing the success of a new emerging idea during the phase of discovery, or participating in a common targeted effort during the phase of consolidation. But for an ambitious young mind the phase of krisis is a unique opportunity. This is the most complex and intense moment of scientific research, when revolutionary and unprejudiced ideas are needed for a real paradigm change.

The phase of krisis in scientific revolutions is epitomised by the birth period of quantum mechanics, when physicists had to knock down the edifice of classical physics and go through a radical paradigm change. Those were ``the thirty years that shook physics" (as Gamow called them~\cite{gamow}) when physicists, in a rebellion against the sacred principles of classical physics, reinvented the map of physical reality. In the words of Abraham Pais: ``Today we live in the midst of upheaval and crisis. We do not know where we are going, nor even where we ought to be going. Awareness is spreading that our future cannot be a straight extension of the past or the present."~\cite{pais1} ``Progress leads to confusion leads to progress and on and on without respite. Every one of the many major advances created sooner or later, more often sooner, new problems. These confusions, never twice the same, are not to be deplored. Rather, those who participate experience them as a privilege."~\cite{pais2}

It is striking how perfectly Pais' words capture the spirit of the intellectual tension that characterises the phase of krisis and how they apply equally well to the times of quantum mechanics and to the present situation. The lack of sense of direction, the confusion on the right track to follow in our quest beyond the Standard Model are not reasons for commiseration but for sensing the privilege of the opportunity for an upcoming paradigm change. While the discovery of the Higgs boson has completed a chapter in the book of science, it has also crystallised new conceptual problems that cry out for solutions. Repeatedly, in the history of science, when knowledge consolidated in a consistent theory, there have been voices announcing that ``there is nothing new to discover." Invariably, they have been proved to be wrong. The chain of ``progress leading to confusion leading to progress" continues without respite and ambitious scientists cherish the moments of confusion.

Of course a big difference between the early 20$^{\rm th}$ century and today is that the paradigm change of quantum mechanics was triggered by many unexpected experimental discoveries, while particle physics today is suffering from a lack of expected experimental discoveries. But the difference may not be so decisive. Although unexpected discoveries are much more sensational and exciting, the lack of expected discoveries can be as effective in triggering a paradigm change. A good example is the failure in measuring the relative motion of the Earth with respect to the aether by Michelson and Morley: a negative result with far-reaching effects! The lack of an expected discovery can be a great source of information able to destroy consolidated principles. What really matters for triggering a paradigm change are two ingredients. One is a broad and intense experimental program at the frontier of knowledge combined with the vitality of a theory community ready to catch in the data the right lead to phrase the correct questions. The other one is the existence of conceptual puzzles regarding fundamental questions. We are lucky to live in an era in which both ingredients are present.

\section{The Post-Naturalness Era}
\label{sec4}
\setlength\epigraphwidth{0.62\textwidth}
\setlength\epigraphrule{0pt}
\renewcommand{\textflush}{flushleft}
\epigraph{\it The ideal is unnatural naturalness, or natural unnaturalness. I mean it is a combination of both.}
{---~~Bruce Lee~\cite{blee}}

I will refer to the present period of rethinking the directions in particle physics as the {\it post-naturalness era}. What are the guiding principles of the post-naturalness era and what are the emerging new paradigms? Of course I have no idea. Today we lack the historical perspective to see through the fog of the present state of healthy confusion. All I can offer are some simple comments (which, most likely, time will show to be wrong) on aspects that could transform the post-naturalness era into a phase of discovery.

\subsubsection*{Naturalness}

Naturalness is a well-defined concept rooted in the logic of effective quantum field theory. If the LHC rules out dynamical solutions to Higgs naturalness at the weak scale, it does not eradicate the problem: a doctor who is unable to find the right diagnosis cannot simply declare the patient healed. Even in post-natural times, the concept of naturalness cannot simply be ignored, although its use may differ from what it has been so far. One way or another, naturalness will still play a role in the post-naturalness era. But, certainly, ruling out dynamical solutions to Higgs naturalness will be one of the most momentous results in particle physics and will have a radical impact on our approach to physics beyond the Standard Model. 

There is already ongoing activity on how the concept of naturalness could be reshaped in post-natural times. Here, instead of giving a comprehensive review, I will only comment on a single new trend: the idea that the explanation of Higgs naturalness may not lie behind some still undiscovered symmetry, but within the cosmological evolution of the universe. 

The most daring approach of this kind is based on a multiverse populated by eternal inflation~\cite{eternal}, in conjunction with the idea that fundamental parameters may not necessarily be god-given numbers, but dynamical variables that take different values in a landscape of vacuum states~\cite{landsc}. While this theoretical setup is quite reasonable, the tricky part is to understand what is the mechanism that singles out our universe, selecting the peculiar values of the cosmological constant and the Higgs vev that we observe in our neighbourhood. There are various possible logical approaches to the selection problem.
One -- the most hated but the most cogent -- is based on anthropic arguments that seem to work reasonably well both for the cosmological constant~\cite{cosmoc} and the Higgs vev~\cite{Agrawal:1997gf}, at least in a context in which only a limited number of parameters scan. The difficulty of this approach is to identify observables that make the hypothesis empirically falsifiable.
Other approaches -- still to be explored, to a large extent -- are based on {\it dynamical selection} (evolution towards a preferred vacuum), {\it statistical selection} (the vast majority of the possible vacua have features similar to our universe), or {\it censorship} (criteria that exclude vacua with the wrong properties).

For me, the greatest mystery behind the multiverse is the diffidence some physicists feel towards it. After all, the idea is part of the established toolkit of theoretical physics. For instance, take the well-respected PQ solution to the strong CP problem~\cite{Peccei:1977hh}. The starting point is the empirical observation that a SM parameter ($\theta_{\rm QCD}$) happens to be smaller than $10^{-10}$ in our universe, while theoretical considerations would give it a value of order one, according to the logic of effective field theory and naturalness. The PQ approach is to promote the SM parameter $\theta_{\rm QCD}$ into a dynamical variable (the axion), so that the actual physical value of $\theta_{\rm QCD}$ in our universe is selected by some underlying dynamics. However, the potential for the axion is generated only after the QCD chiral phase transition, {\it i.e.} very late (at least from a particle theorist's perspective) in the history of the universe. Before that time, the axion (and so the effective $\theta_{\rm QCD}$) takes randomly different values in different parts of space, which are then blown up by inflation into different patches of the universe. In other words, a multiverse is created. Although $\theta_{\rm QCD}$ will eventually relax to zero everywhere, the different patches of the ``multiverse" contain different physical information, because the energy stored in the axion oscillations around its minimum varies from patch to patch, as it depends on the initial condition of the axion. In other words, the axion can be viewed as an incarnation of the multiverse approach. 

The multiverse approach to naturalness is often viewed as the antithesis of the symmetry approach. However, in order to formulate a quantitative criterion of naturalness in the symmetry framework (supersymmetry being the prototypical example) one makes statistical comparisons in a ``theory space", whose coordinates are theory parameters~\cite{Barbieri:1987fn}. Ironically, defining the degree of naturalness of a fundamental theory requires a description in terms of a form of ``multiverse". 

The multiverse is a fertile and constructive framework that could help us to reformulate in different terms some of the open questions in fundamental physics. The idea that the parameters of the Higgs potential or the cosmological constant could be dynamical variables offer a variety of novel approaches to the issue of naturalness. It is intriguing that the Higgs naturalness problem can be rephrased from an issue of UV sensitivity to a problem of criticality~\cite{Giudice:2006sn}: why is the Higgs vev so close to the critical point for electroweak phase transition? From this perspective, it appears that the SM parameters are magically chosen such that the theory lives at the boundary between the broken and unbroken phases. Surprisingly, the discovery of the Higgs boson at 125 GeV has revealed another potential problem of criticality within the SM~\cite{Buttazzo:2013uya}: why is the Higgs mass so close to the critical point for a new phase transition? If taken seriously, these questions may suggest an underlying dynamics of the Higgs potential parameters that drive them towards critical points in the phase diagram. The Higgs potential could undergo a self-tuning process, with critical boundaries acting as attractors, just as in the mechanism of self-organised criticality~\cite{Bak:1987xua}. While such a mechanism of {\it dynamical selection} is uncommon in particle physics, it is rather ubiquitous in other sciences and it has been claimed as an explanation of many disparate phenomena, such as the pattern of earthquakes, forest fires, solar flares, epidemics, and even the behaviour of financial markets.

A concrete realisation of a {\it dynamical selection} mechanism is the relaxion~\cite{Graham:2015cka}. While the Higgs mass parameter is scanned by a slow-rolling field during the evolution of the universe, the electroweak phase transition generates a back-reaction able to stop the evolution near the critical point. An example of {\it statistical selection} is the framework proposed in ref.~\cite{Arvanitaki:2016xds}, where a large number of vacua is generated in the proximity of a special value for electroweak breaking. Only in that region, the cosmological constant can scan finely enough to be within the narrow range of anthropically allowed values.

Much of the popularity of the multiverse was brought about by the discovery of the enormous {\it landscape} of possible vacua in string theory~\cite{landsc}. This result had a negative impact on string theory because it conveyed the feeling of practical ineffectiveness of the UV theory to determine the physical properties of our universe, hampering the enthusiasm towards an ultimate fundamental theory. More recent research has inverted the trend by focusing on the restrictions imposed by the quantum theory of gravity on low-energy effective theories. A large variety of theories which, from the low-energy point of view, look perfectly healthy are actually incompatible with quantum gravity and hence cannot describe physical reality. Using current jargon, these non-theories are said to live in the {\it swampland}, which is believed to be vastly larger than the landscape~\cite{swamp}. This result is a revenge of the ideology professing that the shape of nature is moulded in the UV. 

A concrete example of criteria that exclude theories superficially viable from a low-energy perspective is the weak gravity conjecture~\cite{ArkaniHamed:2006dz}. Although neither the weak gravity conjecture nor other swampland criteria have been proved rigorously, no counterexample in string theory has been found so far. This suggests that an important ingredient for theory selection in the post-naturalness era could be {\it censorship}. UV information may not be sufficient to determine uniquely the final theory, but can effectively prune the tree of possibilities. This line of thought has been pursued in refs.~\cite{Ibanez:2017oqr} and~\cite{Lust:2017wrl} where, using different approaches, it has been suggested that the SM with an electroweak breaking scale much above the TeV scale may only belong to the swampland.

A more prosaic way of implementing {\it censorship} is to invent cosmological criteria that prevent the physical occurrence of sectors with large electroweak breaking scale. One example is  Nnaturalness~\cite{Arkani-Hamed:2016rle}, in which the reheating process acts as a cosmological censor expunging immoral versions of the universe.

Although none of these examples may be a convincing explanation of Higgs naturalness, these models testify to the activity in trying to explain Higgs naturalness as a result of the cosmological evolution. The general approach is an interesting and a novel line of thought. 

While new ideas will emerge in the future, what was learned during the naturalness era will certainly not turn out to be useless for the post-naturalness era. The examples I presented focus on the special features of the Higgs naturalness problem. But it is likely that the Higgs problem is only a small aspect of a bigger naturalness problem. The Higgs boson is unlikely to be the only scalar particle we must reckon with. Other fundamental scalar particles may be required by various patterns of symmetry breaking at high scales, by the axion sector, inflation, and the dark sector. Each of these particles would come with its own naturalness problem. While the Higgs naturalness (or at least the little hierarchy problem) may be cured by alternative medicine, it is plausible that a more radical solution, based on symmetry, is present at the fundamental level and deals with the big hierarchy problem. Supersymmetry is the obvious candidate to reappear in the post-naturalness era in a different guise. One interesting example is split supersymmetry~\cite{split}, which offers many attractive theoretical features, although it lacks the power to deal with the little hierarchy. Even in the context of the relaxion, supersymmetry can help in making the picture more convincing by supplementing UV ingredients lacking in the non-supersymmetric version~\cite{Batell:2015fma}.

\subsubsection*{Cosmological constant}

While discussing the role of naturalness in the post-naturalness era, I focused on the Higgs. However, the cosmological constant poses an even bigger problem. The dominant attitude during the naturalness era was to believe that the two problems could be addressed separately. This was probably not dictated by a deep theoretical conviction, but mostly due to pragmatic reasons. While progress with the cosmological constant was difficult, Higgs naturalness seemed an endless source of revolutionary ideas with promises of direct experimental verification. 

From a post-natural perspective, it seems that the problem of the cosmological constant can no longer be ignored. By making a field expansion of the SM effective potential, it appears that the problem with the cosmological constant and the Higgs mass have a common origin. 
The only difference is that the vacuum energy, unlike the Higgs mass, becomes an observable only through its coupling to gravity. But it seems likely that progress during the post-naturalness era will only come by addressing the two problems simultaneously.

The cosmological constant corresponds to a scale of about $2\times 10^{-3}$~eV and affects physics in the deep IR, at astronomical and cosmological distances. Modifications of gravity at large distances are attempts to tackle the problem from an IR perspective~\cite{deRham:2014zqa}. However, the conceptual problem with the cosmological constant comes from quantum effects in the deep UV. Attempts to tackle the UV problem with the traditional methods used for the Higgs seem hopeless because the corresponding UV cutoff is way lower than any energy scale entering those theories. 

This confusion among scales is at the basis of the problem. It is a big source of confusion because the systematic approach of effective field theories has taught us how to separate energy scales in successive shells and make sense of the theory at each shell separately.  The cosmological constant seems to resist this approach. Naturalness is an offspring of effective field theory and so it is not surprising that the difficulty we are encountering with the effective theory description leads to a problem with naturalness.

A theory that manifests an active interplay between the IR and the UV would not be simply describable by an effective field theory, as it would violate its inner logic. It is not impossible that quantum gravity will exhibit some kind of IR/UV interplay. An indication could come from the classical behaviour of gravity. Consider the head-on collision of two particles at ever increasing energies. Once you pass the threshold for forming a black hole, the more energy you feed in the system the larger the Schwarzschild radius becomes. In other words, higher energy collisions are less sensitive to short distances, in contrast with our effective-theory intuition for a separation between IR and UV.

A radical conclusion that could be derived from these considerations is that we are facing the end of validity of field theory and the solution of the cosmological constant lies beyond our familiar theories. The new framework should incorporate an interplay of physical effects occurring at all scales. Although it is difficult to tell how such a framework would look, one can easily expect that the cosmological constant problem will play a key role in the post-naturalness era.

\subsubsection*{Simplicity and complexity}

The phenomena that we observe in nature are complex. And yet, humans have always had the aspiration to understand nature in terms of an inner simplicity, whether in the form of deities or of physical laws expressed in mathematical language. This idea reverberates also in the world of art. As the sculptor Constantin Br{\^a}ncu{\c{s}}i said, ``Simplicity is complexity resolved."~\cite{branc}

Just by staring at the world outside the window, it may seem that this program of searching for an ultimate physical law governing the entire physical reality is a bit simpleminded. {\it Why would the complexity of natural phenomena be reducible to a simple set of physical laws?} One of the most stupefying results of modern science was to find empirical evidence that this program -- simpleminded as it may seem -- is actually marvellously successful. Nature shows the same aspiration as humans to have simple principles behind the manifest complexity of the universe.\footnote{Probably it is the other way around.} 

Natural phenomena, when observed with poorer resolution, become simpler. This is simple fact. When we describe the Moon trajectory orbiting around Earth, we can safely approximate both bodies with material points and neglect all the complexity of geography, crystalline structure, chemical composition, and terrestrial biology. This is technically known as the ``spherical-cow approximation", which is a basic tool of effective field theory. The observation that the universe is roughly uniform and isotropic at large scales supports the point of view that simplicity dominates the far IR. 

As we move from large to small distances, the complexity of nature grows. We see clusters with thousands of galaxies, each containing billions of star, planets, life. But, surprisingly, once we reach much smaller distances the trend of growing complexity seems to reverse direction. Simplicity emerges again at small distances, in the form of a finite number of building blocks. Chemistry organises matter into molecules, which can be reduced to universal atoms, constructed out of elementary particles. This is a highly non-trivial, and a priori unexpected, result. Simplicity is not relegated to the spherical-cow approximation, but becomes preponderant also at small distances. 

When we enter the particle world, simplicity at small distances takes on a different meaning. As we descend to shorter distances, we discover that physical reality is described not by less building blocks, but more. Heavy quarks, heavy gauge bosons and the Higgs are found only when we explore smaller distances. If supersymmetry were correct, many more particles would be hidden at even smaller distances. We don't know if the SM particles are truly fundamental or will be ultimately replaced by new building blocks. But we have learned that simplicity in the UV does not manifest itself in terms of fewer degrees of freedom, but in terms of more powerful organising principles. The full structure of the fundamental laws becomes apparent only when we observe nature at the smallest possible distance. This is the reason why fundamental physics today is focusing so much on particle physics and high-energy collider experiments. It is there that we expect to find the answers to our most fundamental questions. 

It is not difficult to imagine designing a universe based on simple laws. The problem is that, most likely, such a universe would also exhibit only simple phenomena and would be unable to produce the complexity that characterises life or consciousness. The original question of the reductionist program then takes a more technical form: {\it how can nature get emerging complexity out of simple principles?}

We do not yet have a full picture of how nature overcomes the dichotomy between simple fundamental laws and complex emergent phenomena. But particle physics has made huge progress in this direction and the key words are {\it gauge theory}. Gauge theory is the essential concept out of which the Standard Model is built: a concept that has all the features of a fundamental principle of nature. It is elegant (based on symmetry considerations), robust (no continuous deformations of the theory are generally allowed), and predictive (given the field content, all processes are described by a single coupling constant). In short, it has all the requirements for a physicist to see simplicity in it. 

The magic about gauge theory lies in the richness of its structure and its ability to produce, out of a simple conceptual principle, a great variety of different manifestations. Long-range forces, short-range forces, confinement, dynamical symmetry breaking are all phenomena described by the same principle. The vacuum structure of gauge theory is unbelievably rich, with $\theta$-vacua, instantons, chiral and gluon condensates, all being expressions of the same theory. The phase diagram at finite temperature and density exhibits a variety of new phenomena and states of matter. In short, gauge theory is an exquisite tool to make complexity out of simplicity. 

Is gauge theory the final answer to overcome the simplicity/complexity dichotomy? In spite of the brilliant success of gauge theory, it is likely that post-natural physicists will look elsewhere for new answers. We cannot be sure that the trend of organising principles hiding in the UV will continue beyond the boundaries of present knowledge. The difficulties with the cosmological constant I was alluding to before may already be a sign for the need of a paradigm change and results from the LHC may reinforce this indication. Losing the belief that nature exhibits its fundamental laws in the UV will be a radical change of perspective.

The multiverse already offers new interpretations for merging simplicity with complexity, as it is a perfect example of how nature can have simple physical laws and generate complexity out of the vacuum structure. Many new approaches are starting to be pursued. Gauge theories, as well as Lorentz invariance and the structure of space-time, may be emerging concepts. Techniques from quantum information lead to surprising new insights on the AdS/CFT correspondence and the theory of quantum gravity, opening the door to completely new interpretations of physical reality, as summarised by the slogan ``It from Qubit."~\cite{qubit} This is physics for the post-naturalness era.
  
\subsubsection*{Symmetry}

Symmetry has been a powerful engine that fuelled the rise of the Standard Model and guided most of the attempts to go beyond the Standard Model during the naturalness era. Its instrumental role in physics is undeniable and its influence will be everlasting. But now lend me your ears: I come to bury symmetry, not to praise it.

It is believed that no global symmetries exist in a quantum theory of gravity~\cite{global}. This suggests that global symmetries are not ingredients of the fundamental theory. Global symmetries can only be approximate and accidental. In other words, they are emergent properties at low energy. A simple example is the rotational symmetry that emerges in the spherical-cow approximation. Examples more pertinent to particle physics are baryon and lepton number in the SM, custodial symmetry of the Higgs model in the limit of vanishing hypercharge and quark mass difference, flavour symmetry in the limit of vanishing Yukawa couplings, isospin symmetry in nuclear forces, chiral symmetry in the pion Lagrangian, and many others. These examples show that global symmetry is a useful concept in the IR. It is useful as a classifier in low-energy theories, both in linearly realised versions (establishing selection rules for forbidden local operators) and non-linear versions (characterising the structure of the Lagrangian of light scalar particles). However, in spite of their practical usefulness in the IR,
they are probably inconsequential in the UV, where the truly fundamental theory is expected to live.

The same ``folk theorem" that rules out continuous global symmetries at the fundamental level applies to discrete (non-gauge) symmetries as well. The Standard Model vouches for this assertion since discrete symmetries like C, P, and T can be explained as the result of Lorentz invariance and the structure of the interactions. Whenever possible, the Standard Model breaks maximally C and CP, both in weak and Yukawa interactions. Surprisingly, this is not true for strong interactions, where the topological structure of the gauge theory allows for a large violation of CP, which is not observed in nature. This anomalous behaviour is taken seriously by most physicists, who believe it cannot be just a fluke of strong interactions, but the indication for some new dynamics yet to be discovered ({\it e.g.} the axion).

We are left with gauge symmetry. The problem with gauge symmetry is that it is not a symmetry in the sense of quantum mechanics. A symmetry is the invariance of the Hamiltonian under transformations of quantum states, which are elements of a Hilbert space. Gauge symmetry is not a symmetry because the corresponding transformation does not change the quantum states. Gauge symmetry acts trivially on the Hilbert space and does not relate physically distinct states. A gauge transformation is like a book by James Joyce: it seems that something is going on, but nothing really happens.

Gauge symmetry is the statement that certain degrees of freedom do not exist in the theory. This is why gauge symmetry corresponds only to as a redundancy of the theory description. The non-symmetry nature of gauge symmetry explains why gauge symmetry, unlike global symmetry, cannot be broken by adding local operators to the action: gauge symmetry is exact at all scales. The only way to ``break" gauge symmetry is adding to the theory the missing degrees of freedom, but this operation is not simply a deformation of the theory (as the case of adding local operators to an action with global symmetry) but corresponds to considering an altogether different theory. The non-symmetry nature of gauge symmetry also explains trivially the physical content of the Higgs theorem. For a spontaneously-broken global symmetry, an infinite number of vacuum states are related by the symmetry transformation. This leads to the massless modes dictated by the Goldstone theorem. In a spontaneously-broken gauge symmetry, there is a single physical vacuum and thus there are no massless Goldstones. Gauge symmetry does not provide an exception to the Goldstone theorem, simply because there is no symmetry to start with.

For gauge symmetry, the word `symmetry' is a misnomer, much as `broken' is a misnomer for spontaneously broken symmetry. But as long as the physical meaning is clear, any terminology is acceptable in human language. The important aspect is that the mathematical language of gauge symmetry (both in the linear and non-linear versions) is extremely powerful in physics and permeates the Standard Model, general relativity, and many systems in condensed matter. As the redundancy of degrees of freedom is mathematically described by the same group theory used for quantum symmetries, the use of the word `symmetry' seems particularly forgivable. 

Does this necessarily make gauge symmetry a fundamental element in the UV?  The property of gauge symmetry of being -- by construction -- valid at all energy scales may naively suggest that gauge symmetry must be an ingredient of any UV theory from which the Standard Model and general relativity are derived. On the contrary, many examples have been constructed -- from duality to condensed-matter systems -- where gauge symmetry is not fundamental, but only an emergent property of the effective theory~\cite{Witten:2017hdv}. Gauge symmetry could emerge in the IR, without being present in the UV theory. If this is the case, gauge symmetry is not the key that will unlock the mysteries of nature at the most fundamental level. The concept of symmetry has given much to particle physics, but it could be that it is running out of fuel and that, in the post-naturalness era, new concepts will replace symmetry as guiding principles.

\subsubsection*{Dark matter}

The WIMP miracle is an attractive and fairly generic consequence of the naturalness approach to the Higgs. But, no matter how appealing it is, the WIMP paradigm will have to be replaced, if future experiments give no evidence in support of the idea. 

How is one to proceed without guidance from the WIMP miracle? An often-used strategy is adding a single elementary particle  for the dark matter and, possibly, another particle to mediate a force between the dark and visible sectors. This approach is justified by criteria of ``minimality" or ``simplicity". While I agree that, given our ignorance about the dark sector, this is the first thing to try, I also think that the approach, rather than ``minimal", looks awfully shortsighted. 

We have ample evidence that nature follows a grand scheme. We recognise in the Standard Model features of this grand scheme. Adding one or two particles only for the reason of generating dark matter is not something that nature would do, if she indeed has a grand scheme in mind. It is much more plausible that the dark matter is only the tip of the iceberg of a sector that serves a structural purpose. This is indeed the logic behind traditional WIMP models. Take the prototypical example of supersymmetry: the WIMP does not come alone, but it is one out of many new particles whose purpose is to stabilise the Higgs mass against quantum corrections. 

Once we accept that dark matter is unlikely to manifest itself in ``minimal" ways, the tree of possibilities grows rapidly. 
Post-naturalness dark matter could manifest itself in innumerably different manners. Why should dark matter be a single particle charged under a single force, while matter in the universe comes in a variety of states and chemical compounds?

Both theoretical activity and experimental searches for dark matter are adapting to this growing number of unconventional and multi-component forms of dark matter. 
Tension with the predictions of collisionless dark matter at small scales~\cite{Weinberg:2013aya} can be relaxed if dark matter is self-interacting~\cite{Wandelt:2000ad}. 
Moreover, dark matter could be much lighter than that expected for a WIMP. The extreme case is the proposal of fuzzy dark matter, made of a particle as light as $10^{-22}$~eV~\cite{Hui:2016ltb}. The light dark-matter window, for masses below the GeV range, is not well covered by conventional searches and several new clever techniques have been proposed to explore this region~\cite{Battaglieri:2017aum}. One of the most appealing aspects of these new ideas is that they are employing techniques borrowed from other sectors of science, from atomic physics to condensed matter. This cross-fertilisation between different research fields is especially needed during a phase of krisis.

In unconventional models, one cannot exclude the possibility that dark matter is present in the universe in different forms, from dust to compact objects of astronomical size. The detection of gravitational waves~\cite{Abbott:2016blz} has opened a new field of exploration, providing a new tool for probing the existence of compact objects made of exotic matter~\cite{Giudice:2016zpa}. These measurements have also revitalised~\cite{Bird:2016dcv} the hypothesis that dark matter is not made by new particles, but by a primordial population of black holes~\cite{Hawking:1971ei}.

The field of dark matter is quickly changing and, both on the theoretical and experimental sides, there is growing interest in alternatives to traditional WIMPs. It is likely that the new trend in dark matter will influence the paradigm search during the post-naturalness era and, conversely, that post-natural research will influence dark matter modelling and experimental strategies.

\subsubsection*{The role of experiment}

Physics is a natural science and experimental observation is the essence of the scientific method upon which physics is based. A phase of krisis cannot mature into a paradigm change without the support of an intense, vigorous, and broad experimental program. A reason for my optimism that the present phase of krisis will soon blossom into an era of new knowledge are the exceptional experimental prospects. In every area of fundamental physics, we are witnessing an explosion of innovative projects. A wealth of new experimental results will soon be available, bringing the promise of the spark that will ignite the new paradigm change.

Observational cosmology is moving fast: the B-mode of the CMB is under siege and new missions probing the CMB are already on the horizon; space and ground surveys will make a leap in measurements of the large scale structure and the properties of dark energy. Direct dark matter experiments are expected to reach the neutrino floor and many new techniques are being proposed to explore the light-mass window. Cosmic ray observations can constrain or give a map of the distribution and properties of dark matter particles. Progress in axion and axion-like experiments is advancing rapidly and the hypothesis of dark matter axions will be conclusively tested. Gravitational wave explorations have only started and a new generation of interferometers -- on earth and in space -- will reach stunning precision with sensitivity to a large range of frequencies, thus opening new and still unimagined ways of probing the universe. The neutrino program is advancing on various fronts: exploration of the neutrino mass in $\beta\beta_0$ decay, measurements of neutrino mixing angles and CP-violating phases, ordering of the mass hierarchy, existence of light sterile neutrinos, and neutrino cosmic background. 
Future long- and short-baseline experiments also have the potential of exploring many exotic phenomena beyond the world of neutrinos. B-factories are starting a new campaign of exploration in flavour processes.  Experiments on the muon $g\! - \! 2$, on rare muon processes, and particle EDMs are formidable probes of short-distance physics. Beam dump experiments will explore the dark side of light and feebly-interacting particles. Many table-top experiments probing macroscopic forces and making precision tests of gravity could furnish unexpected surprises.

CERN is contributing to this landscape of experimental activity by promoting studies for possible small-scale experiments that will enrich its scientific program~\cite{PBC}. Of course, the main focus of today's CERN program is the LHC: data taking is in full swing and the operation of the ten-year high-luminosity phase will start in 2026. As of today, the most tantalising LHC news comes from B physics, in the form of combined anomalies in various decay processes~\cite{Blanke:2017qan}. The surprising aspect of these results is that they imply violation of lepton universality, which is not the first place at the high-energy frontier where one would have expected new physics to come up. The exciting aspect of these results is that they involve the flavour sector, which remains mysterious from the theory point of view, as no significant progress has ever been made in computing quark and lepton masses and mixing angles or understanding the dynamical origin of their pattern. The craziness of these results is promising. Given our failure in cracking the flavour problem, it is unlikely that its solution is simple and that it manifests itself in simple ways. Only crazy results in flavour physics have a chance to be true.

While LHC operations are underway, there is already feverish activity around the world in planning the new generation of high-energy collider experiments. The physics community will soon be called upon to make strategical choices about the future of experiments in collider physics. What are the goals of the high-energy physics program in the post-naturalness era?

One sure priority is the research program in Higgs physics. The discovery of the Higgs boson~\cite{Aad:2012tfa} opened the door to the understanding of the mechanism of electroweak breaking. This mechanism plays a crucial role in nature, giving rise to a fundamental scale that rules not only the microscopic world but many physical properties of our universe, from the size of atoms to the timescale of the processes that make the sun shine. Missing the opportunity to study in depth the mechanism of electroweak breaking would be like giving up the exploration of a new continent in the planet of knowledge.

Although a fundamental milestone, the discovery of the Higgs cannot be regarded as the final resolution to the enigma of electroweak breaking. If anything, its discovery has rendered unavoidable the need to address  some of the open problems in particle physics. Indeed, every single Higgs interaction introduces its own puzzle. The Yukawa couplings express the flavour problem, {\it i.e.} our inability to compute quark and lepton masses and mixing angles. The Higgs quartic coupling raises issues with the stability of the SM electroweak vacuum. The Higgs quadratic term incarnates the naturalness problem. The constant term in the Higgs potential is an expression of the cosmological constant problem.

The root of these problems is that the Higgs boson introduces 14 new forces (without taking into account other forces associated with neutrinos) besides the 3 known fundamental forces of the Standard Model. Unlike strong, weak, and electromagnetic forces, these new 14 forces are not gauge-like. Not surprisingly, they do not share the properties of elegance, simplicity, robustness, and predictivity that characterise the gauge forces of the Standard Model. Accustomed with the conceptual perfection of the gauge sector of the Standard Model, it is difficult for a theorist to believe that a structure so arbitrary and provisional as the Higgs sector could be the final word on electroweak breaking. There must be more still to be discovered.

The way to get to the bottom of the question is a program of precision measurements on the Higgs boson properties. This is already underway, but future high-energy colliders will be able to bring these precision studies to unprecedented levels of precision. Much can be learned from these studies about the Higgs boson, the phenomenon of electroweak breaking, the fundamental laws that govern nature, the early stages of the Universe and its ultimate fate.
 
The recent history of experimental particle physics has been characterised by large projects that had the exploration of the unknown as primary goal but, at the same time, had also a clear objective within reach. It was the discovery of the $W$ and $Z$ for the SPS, the top for Tevatron, precision studies of the gauge sector for LEP, the Higgs for LHC. However, the story may change in the future and experimental physics during the post-naturalness era must reckon with situations in which clear objectives within reach are difficult to be determined. Any result from the LHC or other experiments can suddenly drive a change in directions and in experimental priorities. Krisis is a phase dominated by uncertainty. But this makes research only more exciting: a daring program of exploration with ambitious goals is what the post-naturalness era needs.
  
Today we are facing many big open problems in the understanding of the particle world, but we are uncertain about where their solutions lie. In the present situation, exploration is the only possible answer to our thirst for knowledge. There is nothing new in this. The very soul of particle physics has always been exploration. Our history is a history of pushing frontiers, crossing boundaries between the known and unknown, exploring virgin territories. But it has never been a blind exploration. We don't pursue a direction just because we have the technology to do so. We do it because we believe that there is something fundamental to learn by pursuing that direction. The measurements that matter are those from which we learn something that we didn't know before. The fantastic success of particle physics has been the result  of wise choices about the best frontiers to push, about the right barriers to break. This success was possible only because of the right mixture of theoretical and experimental physics, of daring ideas and fabulous technologies. 

Our scientific priorities are likely to shift in the post-naturalness era, but the high-energy frontier seems, more than ever, the most promising direction for us to gain new knowledge in fundamental physics. In the post-naturalness era we will have to tackle new questions, both theoretically and experimentally. All indications point towards higher energy as the right frontier to push in order to answer many fundamental questions about our universe. This is what we can do with future high-energy collider projects. 

\subsubsection*{The role of theory}

``The truth is, the Science of Nature has been already too long made only a work of the Brain and the Fancy: It is now high time that it should return to the plainness and soundness of Observations on material and obvious things." These words are not the latest attack against the abstractness of modern theoretical physics or the inability of string theory to find a criterion for empirical falsifiability. No, these words are by Robert Hooke and date 1665~\cite{hooke}, when times were changing in England and a new approach to science was emerging: a radically different use of abstract mathematics in formulating physical theories. Only one year later, in the {\it annus mirabilis} 1666, Isaac Newton, who had no sympathy for Hooke -- sincerely reciprocated by Hooke -- started to revolutionise physics by making use of calculus in his formulation of mechanics and gravitation. A masterpiece of a work of the Brain.

The criticism for excessive mathematical abstractness has accompanied theoretical physics throughout all its major breakthroughs. It was the case for Maxwell's highly mathematical treatise {\it A Dynamical Theory of the Electromagnetic Field}, which gives the foundation of modern electromagnetism and predicts electromagnetic waves travelling at the speed of light. ``For more than twenty years, his theory of electromagnetism was largely ignored," recounts Dyson. ``It was regarded as an obscure speculation without much experimental evidence to support it."~\cite{dyson} Then came Einstein's relativity theory, which was rejected by many, including the Nobel laureate Philipp Lenard, for ``lacking the illustrativeness of classical physics''~\cite{lenard1} and, in later darker times, for contradicting the principles of ``Aryan physics by man of Nordic kind [\dots ] which only accepts scientific knowledge based on experiments, and only if accessible to the senses."~\cite{lenard2} At the beginning of last century, the abstract developments of quantum mechanics -- from Schr\"odinger's complex analysis describing wavefunctions to Heisenberg's non-commutative operator algebra -- were targets of innumerable grievances about how theoretical physics had departed from physical reality. Today similar lamentations about the status of theoretical physics are not uncommon.

The role of logical deduction in physics and the need to develop a mathematical language for the description of nature have always been clear to the great minds of our field, starting from Galileo and Newton at the very beginning of modern science and continuing throughout the history of physics. The concept was expressed by Einstein during a lecture at Oxford in 1933: ``It is my conviction that pure mathematical construction enables us to discover the concepts and the laws connecting them which give us the key to the understanding of the phenomena of Nature. Experience can of course guide us in our choice of serviceable mathematical concepts; it cannot possibly be the source from which they are derived; experience of course remains the sole criterion of the serviceability of a mathematical construction for physics, but the truly creative principle resides in mathematics. In a certain sense, therefore, I hold it to be true that pure thought is competent to comprehend the real, as the ancients dreamed."~\cite{einstein1}

Dirac pushed the concept further, elaborating on the creative role of theoretical physics in employing mathematics: ``The steady progress of physics requires for its theoretical formulation a mathematics that gets continually more advanced. This is only natural and to be expected. What, however, was not expected by the scientific workers of the last century was the particular form that the line of advancement of the mathematics would take, namely, it was expected that the mathematics would get more and more complicated, but would rest on a permanent basis of axioms and definitions, while actually the modern physical developments have required a mathematics that continually shifts its foundations and gets more abstract. Non-euclidean geometry and non-commutative algebra, which were at one time considered to be purely fictions of the mind and pastimes for logical thinkers, have now been found to be very necessary for the description of general facts of the physical world. It seems likely that this process of increasing abstraction will continue in the future and that advance in physics is to be associated with a continual modification and generalisation of the axioms at the base of mathematics rather than with a logical development of any one mathematical scheme on a fixed foundation."~\cite{dirac2}

Physics doesn't simply borrow the language from mathematics, but manipulates it, adapts it, and reinvents it according to the needs for a description of physical reality. The magic about physics is transforming the pure logic of mathematics into a beautiful narrative about nature. A poet couldn't write poetry without a language, but language is not sufficient to make poetry. The same happens in science: mathematics is the language, but it takes physics to do poetry.

I leave the punchline to Dirac: ``I think there is a moral to this story, namely that it is more important to have beauty in one's equations than to have them fit experiment."~\cite{dirac3}

It is Popperianly evident that empirical confirmation is the ultimate judge on any physical theory; only a deranged scientist would deny this truth. It is also clear that experiments are indispensable in guiding theorists to ask the right questions. But denying the speculative nature of physics is betraying the very spirit of our discipline. 
Logical deduction is the engine that has always driven physicists to discern the inner structure of physical reality and the laws that govern the universe. Most of the sparks that ignited major advances in physics can be traced back to pure thought, and not to the motivation of explaining a certain observation. It is difficult to imagine how Einstein could have discovered general relativity, had he started by trying to fix the observed discrepancy in the precession of Mercury's perihelion.\footnote{If one takes the discrepancy in Mercury's data as starting point, the most straightforward solution would be postulating a new planet causing a perturbation in Mercury's orbit. This is the approach followed by Urbain Le Verrier, who had been previously successful in predicting the existence of Neptune, and who tried his luck with a new planet he named Vulcan. The discovery of Vulcan was announced by Le Verrier in 1860, on the basis of observations made by an amateur astronomer, Edmond Modeste Lescarbault. The debate over the observations lasted for decades and, when he died in 1877, Le Verrier was still convinced about Vulcan's existence~\cite{vulcan}.} Logical deduction, based on arguments like the equivalence principle, is what brought Einstein to his monumental theoretical construction. 

While it is always good to keep in mind the role of experiments as inspirational guideline and ultimate arbiter, the call to abandon ``the Brain and the Fancy" and return ``to the plainness and soundness of Observations on obvious things" can have destructive consequences. In times of krisis, more than ever, we need pure unbridled speculation, driven by imagination and vision. Paradigm changes require unbiased thinking and the freedom to blaze new trails. Support for an intense and diverse programme in theoretical physics -- from the most speculative areas to those closely-connected with experiment -- is the central element for success in the post-naturalness era.

Exchange and communication between neighbouring fields is also an essential fertiliser for theoretical physics. When in search for new directions, one must look beyond the boundaries of a single discipline. The transfer of ideas is a key to breakthroughs. A good historical example is the concept of spontaneously broken symmetry, which has so radically influenced particle physics -- from pion interactions to the Standard Model -- but was pioneered in condensed matter physics. Today, at the dawn of the post-naturalness theory, we start observing a revival in the transfer of ideas, not only with fields traditionally akin to particle physics -- such as mathematics, astrophysics and cosmology -- but with other areas as well. Ideas originally formulated in condensed matter, quantum information, and atomic physics are finding unexpected applications in particle physics (examples are the use of tensor networks in holography and quantum gravity~\cite{tensor}, and the quantum simulation of field theories with ultracold atoms or photonic systems~\cite{cirac}). At the same time, the concept of the AdS/CFT correspondence~\cite{Maldacena:1997re} developed from string theory is finding its way into condensed matter~\cite{Sachdev:2010ch} and heavy-ion physics~\cite{CasalderreySolana:2011us}.

What matters is not the difference between ``abstract" and ``real" physics, but between ``good" and ``bad" physics: the difference between physics that is imaginative, original, ambitious in the pursuit of fundamental questions; and physics that is misguided, ill grounded, or simply repetitive and unimaginative. Of course, the distinction between the two is not always easy to discern at the beginning and, since research will never be an efficient process, a large number of unproductive attempts and false starts must be expected and taken into account. But this cannot be used as an argument to accept lower scientific standards. On the contrary, logical rigour and verifiable truth remain the ultimate guiding principles to discriminate between ``good" and ``bad" physics. 

Theoretical physics is the driving force needed in the search for a new paradigm during the post-naturalness era.
The aim for our community is to create an environment that supports and encourages free thinking in theoretical physics, preparing a fertile ground able to nurture the single idea that will eventually become the paradigm changer. 

\section{Conclusions}
\setlength\epigraphwidth{.84\textwidth}
\setlength\epigraphrule{0pt}
\renewcommand{\textflush}{flushepinormal}
\epigraph{\it It was the best of times, it was the worst of times, it was the age of wisdom, it was the age of foolishness, it was the epoch of belief, it was the epoch of incredulity, it was the season of Light, it was the season of Darkness, it was the spring of hope, it was the winter of despair, we had everything before us, we had nothing before us, we were all going direct to Heaven, we were all going direct the other way.}
{---~~Charles Dickens~\cite{dickens}}

When I started studying physics, I fell in love with the era when quantum mechanics was born. An era of confusion, inexplicable experimental results and crazy theoretical ideas, rebellion against the sacred laws of physics and heated debates at the Solvay conferences. How much I wished I could have been there: I missed my chance, I thought.

As I advanced in my studies, I was captivated by the excitement of the post-war period, when physicists had to deal with untamable infinities and a zoo of particles, produced by cosmic rays or popping out from pioneering accelerators, and whose meaning was obscure. The era of Shelter Island and the birth of QED, the attempts with Regge trajectories, S-matrix, bootstrap to end up with gauge quantum field theory. How much I wished I could have been there: I missed my chance, I thought.

Well, now I realise that my chance is today. As a scientist, I have the privilege to live in a new era of krisis. Ideas thrive
in the periods of krisis dominated by uncertainty and confusion, when physicists are in search of a paradigm change able to deal with the puzzles they are confronted with. There is no lack of open fundamental questions we must tackle today: the nature of the Higgs boson, the structure of quarks and leptons, inflation, the cosmic baryon asymmetry, dark matter, dark energy, quantum gravity, and more. But there is also a widespread feeling that our theoretical tools -- which have been so successful in bringing particle physics to its present stage of maturity -- are becoming inadequate to address the next layer of open questions. A new paradigm change seems to be necessary.

Experimental physics is reacting to the present status of krisis with a broad and ambitious program that will enable humanity to cross the borders of knowledge on many fronts. Theoretical physics is exploring new directions and looking beyond the boundaries of traditional particle physics, across different disciplines. Revolutions in science don't happen overnight: it took thirty years for quantum mechanics to develop from Planck's black-body radiation to Dirac's equation; twenty-five years for the Standard Model to go from QED to the asymptotic freedom of QCD. We can't expect to find all answers today. But we are experiencing all the right symptoms -- unresolved fundamental questions, an old paradigm that seems to run out of mileage, bold experimental projects, revived theoretical curiosity -- that indicate we are living in the dawn of a new era. This is the post-naturalness era, which may soon become a new chapter in the continuing story of human exploration of knowledge.

\bigskip

\noindent{\bf Acknowledgements}

I thank Nima Arkani-Hamed, Eckhard Elsen, Debra Giudice, Michelangelo Mangano, Matthew McCullough, and Riccardo Rattazzi for comments on the manuscript. I thank all my friends and colleagues for the many discussions that continuously reshape my views on particle physics. And I thank Stefano Forte, Aharon Levy, and Giovanni Ridolfi for taking the initiative of creating this tribute to the memory of Guido Altarelli.


\footnotesize 
\begin{thebibliography}{20}
\parskip=0pt
\baselineskip=12pt

\bibitem{tibet}
{\it A Collection of Tibetan Proverbs and Sayings: Gems of Tibetan Wisdom and Wit}, edited by C.~C\"uppers and P.~K.~S{\o}rensen, Franz Steiner Verlag, Stuttgart 1998.

\bibitem{altwork}
G. Altarelli, Summary Talk at {\it International Workshop on Future Hadron Colliders: Physics, Detectors, Machines}, Fermilab, 16-18 October 2003;\\
{\scriptsize \tt <http://conferences.fnal.gov/hadroncollider/talks/Altarelli.pdf>}.

\bibitem{Altarelli:1997ce}
  G.~Altarelli, J.~R.~Ellis, G.~F.~Giudice, S.~Lola and M.~L.~Mangano,
  {\it Pursuing interpretations of the HERA large $Q^{2}$ data},
  Nucl.\ Phys.\ B 506 (1997) 3
  [arXiv:hep-ph/9703276].

\bibitem{nat}
F.-J.~de Pierre, Cardinal de Bernis, {\it R\'eflexions sur les passions et sur les go\^uts}, Didot, Paris 1741, as translated in {\it Dictionary of Quotations (French and Italian)}, edited by T.~B.~Harbottle and P.~H.~Dalbiac, Swan Sonnenschein \& Co., London 1904.

\bibitem{Giudice:2008bi}
  G.~F.~Giudice,
  {\it Naturally Speaking: The Naturalness Criterion and Physics at the LHC}, in {\it Perspectives on LHC physics}, edited by G.~Kane and A.~Pierce, World Scientific, Singapore 2008 [arXiv:0801.2562].

\bibitem{Giudice:2013nak}
  G.~F.~Giudice,
  {\it Naturalness after LHC8},
  PoS EPS-HEP2013 (2013) 163
  [arXiv:1307.7879].

\bibitem{Aad:2012tfa}
  G.~Aad {\it et al.} [ATLAS Collaboration],
  {\it Observation of a new particle in the search for the Standard Model Higgs boson with the ATLAS detector at the LHC},
  Phys.\ Lett.\ B 716 (2012) 1
  [arXiv:1207.7214].
  S.~Chatrchyan {\it et al.} [CMS Collaboration],
  {\it Observation of a new boson at a mass of 125 GeV with the CMS experiment at the LHC},
  Phys.\ Lett.\ B 716 (2012) 30
  [arXiv:1207.7235].

\bibitem{donof}
M.~D'Onofrio (on behalf of ATLAS, CMS, and LHCb Collaborations), {\it Supersymmetry and Exotics Searches}, talk at the EPS Conference on High Energy Physics, Venice 5-12 July 2017.

\bibitem{dalai}
{\it The Dalai Lama Book of Quotes: A Collection of Speeches, Quotations, Essays and Advice from His Holiness}, edited by T. Hellstrom, Hatherleigh Press, Hobart 2016.

\bibitem{kuhn}
T.~S.~Kuhn, {\it The Structure of Scientific Revolutions}, University of Chicago Press, Chicago 1962.

\bibitem{kennedy1} 
J.~F.~Kennedy, Remarks at the United Negro College Fund, Indianapolis, 12 April 1959; Box 902, Senate Speech Files, Pre-Presidential Papers, John F. Kennedy Presidential Library.

\bibitem{zimmer} 
B.~Zimmer, {\it Crisis = danger + opportunity: The plot thickens}, Language Log, 27 March 2007.

\bibitem{kennedy2} 
J.~F.~Kennedy, Television report to the American people on civil rights, 11 June 1963.

\bibitem{dante}
Dante Alighieri, {\it Divina Commedia}, Inferno III, vv. 34-36.

\bibitem{gamow}
G.~Gamow, {\it Thirty Years That Shook Physics: The Story of Quantum Theory}, Anchor Books, Garden City 1966.

\bibitem{pais1}
A.~Pais, {\it Physics in Denmark: The First Four Hundred Years}, 75th Anniversary of the Niels Bohr Institute, 6 March 1996.

\bibitem{pais2}
A.~Pais, {\it Inward Bound. Of Matter and Forces in the Physical World}, Oxford University Press, Oxford 1986.

\bibitem{blee}
B.~Lee and J.~R.~Little, {\it Bruce Lee: Artist of Life}, Tuttle Publishing, Boston 1999.

\bibitem{eternal} 
   A.~Vilenkin,
  {\it The Birth of Inflationary Universes},
  Phys.\ Rev.\ D 27 (1983) 2848.
  A.~D.~Linde,
  {\it Eternal Chaotic Inflation},
  Mod.\ Phys.\ Lett.\ A 1 (1986) 81.
    A.~D.~Linde,
  {\it Eternally Existing Selfreproducing Chaotic Inflationary Universe},
  Phys.\ Lett.\ B 175 (1986) 395.
  
\bibitem{landsc} 
  R.~Bousso and J.~Polchinski,
  {\it Quantization of four form fluxes and dynamical neutralization of the cosmological constant},
  JHEP 0006 (2000) 006
  [arXiv:hep-th/0004134].
  S.~Kachru, R.~Kallosh, A.~D.~Linde and S.~P.~Trivedi,
  {\it De Sitter vacua in string theory},
  Phys.\ Rev.\ D 68 (2003) 046005
   [arXiv:hep-th/0301240].
   L.~Susskind,
  {\it The Anthropic landscape of string theory}, in {\it Universe or multiverse?}, edited by B.~Carr, Cambridge University Press, Cambridge 2007
  [arXiv:hep-th/0302219].
    M.~R.~Douglas,
  {\it The Statistics of string / M theory vacua},
  JHEP 0305 (2003) 046
  [arXiv:hep-th/0303194].
 L.~Susskind,
{\it The Cosmic Landscape: String Theory and the Illusion of Intelligent Design}, Little, Brown and Company, New York 2005.
 
\bibitem{cosmoc} 
  S.~Weinberg,
  {\it Anthropic Bound on the Cosmological Constant},
  Phys.\ Rev.\ Lett.\  59 (1987) 2607.
  
\bibitem{Agrawal:1997gf}
  V.~Agrawal, S.~M.~Barr, J.~F.~Donoghue and D.~Seckel,
  {\it The Anthropic principle and the mass scale of the standard model},
  Phys.\ Rev.\ D 57 (1998) 5480
  [arXiv:hep-ph/9707380].

\bibitem{Peccei:1977hh}
  R.~D.~Peccei and H.~R.~Quinn,
  {\it CP Conservation in the Presence of Instantons},
  Phys.\ Rev.\ Lett.\  38 (1977) 1440.

\bibitem{Barbieri:1987fn}
  R.~Barbieri and G.~F.~Giudice,
  {\it Upper Bounds on Supersymmetric Particle Masses},
  Nucl.\ Phys.\ B 306 (1988) 63.

\bibitem{Giudice:2006sn}
  G.~F.~Giudice and R.~Rattazzi,
  {\it Living Dangerously with Low-Energy Supersymmetry},
  Nucl.\ Phys.\ B 757 (2006) 19
  [arXiv:hep-ph/0606105].

\bibitem{Buttazzo:2013uya}
  D.~Buttazzo, G.~Degrassi, P.~P.~Giardino, G.~F.~Giudice, F.~Sala, A.~Salvio and A.~Strumia,
  {\it Investigating the near-criticality of the Higgs boson},
  JHEP 1312 (2013) 089
  [arXiv:1307.3536].

\bibitem{Bak:1987xua}
  P.~Bak, C.~Tang and K.~Wiesenfeld,
  {\it Self-organized criticality: An Explanation of 1/f noise},
  Phys.\ Rev.\ Lett.\  59 (1987) 381.
 
 \bibitem{Graham:2015cka}
  P.~W.~Graham, D.~E.~Kaplan and S.~Rajendran,
  {\it Cosmological Relaxation of the Electroweak Scale}, Phys.\ Rev.\ Lett.\  115 (2015) 221801 [arXiv:1504.07551].

\bibitem{Arvanitaki:2016xds}
  A.~Arvanitaki, S.~Dimopoulos, V.~Gorbenko, J.~Huang and K.~Tilburg,
  {\it A small weak scale from a small cosmological constant}, JHEP 1705 (2017) 071 [arXiv:1609.06320].

\bibitem{swamp}
  C.~Vafa,
  {\it The String landscape and the swampland}, arXiv:hep-th/0509212.
  H.~Ooguri and C.~Vafa,
  {\it On the Geometry of the String Landscape and the Swampland},
  Nucl.\ Phys.\ B 766 (2007) 21
  [arXiv:hep-th/0605264].

\bibitem{ArkaniHamed:2006dz}
  N.~Arkani-Hamed, L.~Motl, A.~Nicolis and C.~Vafa,
  {\it The String landscape, black holes and gravity as the weakest force}, JHEP 0706 (2007) 060
  [arXiv:hep-th/0601001].

\bibitem{Ibanez:2017oqr}
  L.~E.~Ibanez, V.~Martin-Lozano and I.~Valenzuela,
  {\it Constraining the EW Hierarchy from the Weak Gravity Conjecture},
  arXiv:1707.05811.

\bibitem{Lust:2017wrl}
  D.~Lust and E.~Palti,
  {\it Scalar Fields, Hierarchical UV/IR Mixing and The Weak Gravity Conjecture},
  arXiv:1709.01790.

\bibitem{Arkani-Hamed:2016rle}
  N.~Arkani-Hamed, T.~Cohen, R.~T.~D'Agnolo, A.~Hook, H.~D.~Kim and D.~Pinner,
  {\it Solving the Hierarchy Problem at Reheating with a Large Number of Degrees of Freedom},
  Phys.\ Rev.\ Lett.\  117 (2016) 251801
  [arXiv:1607.06821].

\bibitem{split}
  N.~Arkani-Hamed and S.~Dimopoulos,
  {\it Supersymmetric unification without low energy supersymmetry and signatures for fine-tuning at the LHC},
  JHEP 0506 (2005) 073
  [arXiv:hep-th/0405159].
  G.~F.~Giudice and A.~Romanino,
  {\it Split supersymmetry},
  Nucl.\ Phys.\ B 699 (2004) 65
  [arXiv:hep-ph/0406088].
  N.~Arkani-Hamed, S.~Dimopoulos, G.~F.~Giudice and A.~Romanino,
  {\it Aspects of split supersymmetry},
  Nucl.\ Phys.\ B 709 (2005) 3
  [arXiv:hep-ph/0409232].
 
\bibitem{Batell:2015fma}
  B.~Batell, G.~F.~Giudice and M.~McCullough,
  {\it Natural Heavy Supersymmetry},
  JHEP 1512 (2015) 162
  [arXiv:1509.00834].
 
 \bibitem{deRham:2014zqa}
  C.~de Rham,
  {\it Massive Gravity},
  Living Rev.\ Rel.\  17 (2014) 7
  [arXiv:1401.4173].

\bibitem{branc}
C.~Gimenez and M.~Gale, {\it Constantin Brancusi: The Essence of Things}, Tate Gallery Publishing, London 2004.
 
 \bibitem{qubit}
{\it It from Qubit: Simons Collaboration on Quantum Fields, Gravity and Information}, Simons Foundation, Mathematical and Physical Sciences.

\bibitem{global}
  R.~Kallosh, A.~D.~Linde, D.~A.~Linde and L.~Susskind,
  {\it Gravity and global symmetries},
  Phys.\ Rev.\ D 52 (1995) 912
  [arXiv:hep-th/9502069].
  T.~Banks and N.~Seiberg,
  {\it Symmetries and Strings in Field Theory and Gravity},
  Phys.\ Rev.\ D 83 (2011) 084019
  [arXiv:1011.5120].

\bibitem{Witten:2017hdv}
  E.~Witten,
  {\it Symmetry and Emergence},
  arXiv:1710.01791.

\bibitem{Weinberg:2013aya}
  D.~H.~Weinberg, J.~S.~Bullock, F.~Governato, R.~Kuzio de Naray and A.~H.~G.~Peter,
  {\it Cold dark matter: controversies on small scales},
  Proc.\ Nat.\ Acad.\ Sci.\  112 (2014) 12249
  [arXiv:1306.0913].

\bibitem{Wandelt:2000ad}
  B.~D.~Wandelt, R.~Dave, G.~R.~Farrar, P.~C.~McGuire, D.~N.~Spergel and P.~J.~Steinhardt,
  {\it Selfinteracting dark matter},
  arXiv:astro-ph/0006344.

\bibitem{Hui:2016ltb}
  L.~Hui, J.~P.~Ostriker, S.~Tremaine and E.~Witten,
  {\it Ultralight scalars as cosmological dark matter},
  Phys.\ Rev.\ D 95 (2017) 043541
  [arXiv:1610.08297].

\bibitem{Battaglieri:2017aum}
  M.~Battaglieri {\it et al.},
  {\it US Cosmic Visions: New Ideas in Dark Matter 2017: Community Report},
  arXiv:1707.04591.

\bibitem{Abbott:2016blz}
  B.~P.~Abbott {\it et al.} [LIGO Scientific and Virgo Collaborations],
  {\it Observation of Gravitational Waves from a Binary Black Hole Merger},
  Phys.\ Rev.\ Lett.\  116 (2016) 061102
  [arXiv:1602.03837].

\bibitem{Giudice:2016zpa}
  G.~F.~Giudice, M.~McCullough and A.~Urbano,
  {\it Hunting for Dark Particles with Gravitational Waves},
  JCAP 1610 (2016) 001
  [arXiv:1605.01209].
V.~Cardoso, S.~Hopper, C.~F.~B.~Macedo, C.~Palenzuela and P.~Pani,
  {\it Gravitational-wave signatures of exotic compact objects and of quantum corrections at the horizon scale},
  Phys.\ Rev.\ D 94 (2016) 084031
  [arXiv:1608.08637].

\bibitem{Bird:2016dcv}
  S.~Bird, I.~Cholis, J.~B.~Mu\~noz, Y.~Ali-Ha\"imoud, M.~Kamionkowski, E.~D.~Kovetz, A.~Raccanelli and A.~G.~Riess,
  {\it Did LIGO detect dark matter?},
  Phys.\ Rev.\ Lett.\  116 (2016) 201301
  [arXiv:1603.00464].
M.~Sasaki, T.~Suyama, T.~Tanaka and S.~Yokoyama,
{\it Primordial Black Hole Scenario for the Gravitational-Wave Event GW150914},
Phys.\ Rev.\ Lett.\  117 (2016) 061101
[arXiv:1603.08338].
  S.~Clesse and J.~Garc\'ia-Bellido,
  {\it The clustering of massive Primordial Black Holes as Dark Matter: measuring their mass distribution with Advanced LIGO},
  Phys.\ Dark Univ.\  15 (2017) 142
  [arXiv:1603.05234].
  A.~Kashlinsky,
  {\it LIGO gravitational wave detection, primordial black holes and the near-IR cosmic infrared background anisotropies},
  Astrophys.\ J.\  823 (2016) L25
  [arXiv:1605.04023].

\bibitem{Hawking:1971ei}
  Y.~B.~Zel'dovich and I.~D.~Novikov, 
{\it The hypothesis of cores retarded during expansion and the hot cosmological model},
Soviet Astronomy 10 (1967) 602.
  S.~Hawking,
  {\it Gravitationally collapsed objects of very low mass},
  Mon.\ Not.\ Roy.\ Astron.\ Soc.\  152 (1971) 75.

\bibitem{PBC}
Physics Beyond Colliders, CERN Working Group; {\scriptsize \tt <https://indico.cern.ch/category/7885/>}.

\bibitem{Blanke:2017qan}
  M.~Blanke,
  {\it Quo vadis flavour physics? - FPCP2017 theory summary and outlook},
  PoS FPCP 2017 (2017) 042
  [arXiv:1708.06326].

\bibitem{hooke}
R.~Hooke, {\it Micrographia: or, Some physiological descriptions of minute bodies made by magnifying glasses}, J.~Martyn and J.~Allestry, London 1665.

\bibitem{dyson}
F.~J.~Dyson, {\it Why is Maxwell's Theory so hard to understand?}, Proceedings of 2nd European Conference on Antennas and Propagation (EuCAP 2007), Edinburgh 2007.

\bibitem{lenard1}
P.~Lenard in {\it Vortr\"age und Diskussionen von der 86. Naturforscherversammlung in Nauheim vom 19.-25. September 1920}, Physikalische Zeitschrift 21 (1920) 649.

\bibitem{lenard2}
P.~Lenard, {\it Deutsche Physik in vier B\"anden}, J.F. Lehmann, Munich 1936.

\bibitem{einstein1}
A.~Einstein, {\it On the Method of Theoretical Physics}, Herbert Spencer lecture delivered at Oxford on 10 June 1933, published in {\it Mein Weltbild}, Querido Verlag, Amsterdam 1934 and in {\it Philosophy of Science}, Vol.~1, No.~2 (1934), p.~163.

\bibitem{dirac2}
  P.~A.~M.~Dirac,
  {\it Quantized Singularities in the Electromagnetic Field},
  Proc.\ Roy.\ Soc.\ Lond.\ A 133 (1931) 60.

\bibitem{dirac3}
 P.~A.~M.~Dirac,
 {\it The Evolution of the Physicist's Picture of Nature},
 Scientific American, Vol. 208, No. 5, May 1963.
 
\bibitem{vulcan}
R.~Baum and W.~Sheehan, {\it In Search of Planet Vulcan: The Ghost in Newton's Clockwork Machine}, Plenum Press, New York 1997. 

\bibitem{tensor}
See, {\it e.g.}, {\it Tensor Networks Initiative}, Perimeter Institute;\\
{\scriptsize \tt <https://www.perimeterinstitute.ca/research/research-initiatives/tensor-networks-initiative>}.

\bibitem{cirac}
J.~I.~Cirac and P.~Zoller, {\it Goals and opportunities in quantum simulation}, Nature Physics 8 (2012) 264.

\bibitem{Maldacena:1997re}
  J.~M.~Maldacena,
  {\it The Large N limit of superconformal field theories and supergravity},
  Int.\ J.\ Theor.\ Phys.\  38 (1999) 1113
   [Adv.\ Theor.\ Math.\ Phys.\  2 (1998) 231]
  [arXiv:hep-th/9711200].
J.~M.~Maldacena,
{\it The illusion of gravity}, Scientific American 293 (2005) 56.

\bibitem{Sachdev:2010ch}
  S.~Sachdev,
  {\it Condensed Matter and AdS/CFT},
  Lect.\ Notes Phys.\  828 (2011) 273
  [arXiv:arXiv:1002.2947].
 S.~Sachdev, {\it Strange and stringy}, Scientific American 308 (2013) 44.
 
\bibitem{CasalderreySolana:2011us}
  J.~Casalderrey-Solana, H.~Liu, D.~Mateos, K.~Rajagopal and U.~A.~Wiedemann,
  {\it Gauge/String Duality, Hot QCD and Heavy Ion Collisions}, Cambridge University Press, Cambridge 2014.
   [arXiv:1101.0618].
  
\bibitem{dickens}
 C.~Dickens, {\it A Tale of Two Cities}, Chapman \& Hall, London 1859.
 
\end{thebibliography}
\end{document}